\documentstyle[aps,prl,multicol,epsf]{revtex}

\begin{document}

\title{Interlayer tunneling spectroscopy of Bi$_2$Sr$_2$CaCu$_2$O$_{8+\delta}$:
\newline a look from inside on the doping phase diagram of high $T_c$ superconductors.}

\author{V.M.Krasnov}

\address{Department of Microelectronics and Nanoscience,
 Chalmers University of Technology, S-41296 G\"oteborg, Sweden}

\date{\today }
\maketitle

\begin{abstract}

A systematic, doping dependent interlayer tunneling spectroscopy
of Bi2212 high $T_c$ superconductor is presented. An improved
resolution made it possible to simultaneously trace the
superconducting gap (SG) and the normal state pseudo-gap (PG) in a
close vicinity of $T_c$ and to analyze closing of the PG at $T^*$.
The obtained doping phase diagram exhibits a critical doping point
for appearance of the PG and a characteristic crossing of the SG
and the PG close to the optimal doping. This points towards
coexistence of two different and competing order parameters in
Bi2212. Experimental data indicate that the SG can form a combined
(large) gap with the PG at $T<T_c$ and that the interlayer
tunneling becomes progressively incoherent with decreasing doping.


\end{abstract}

\begin{multicols}{2}

Observation of an energy gap in the electronic density of states
(DOS) had a decisive role in understanding of low $T_c$
superconductivity \cite{Giaevier}. However, fifteen years after
discovery of high $T_c$ superconductors (HTSC), there is still no
consensus about HTSC energy gap. Several experiments revealed
different energy scales in HTSC
\cite{Deut,KrTemp,Campuz,Puchkov,Tallon,KrMag}. One of those, a
normal state pseudo-gap (PG), persists at $T>T_c$
\cite{Deut,KrTemp,Campuz,Puchkov,Tallon,KrMag,Fisher}. The origin
of the PG is an intriguing open question, which is crucial for
understanding HTSC. Currently, the scientific community is
divided, believing either in superconducting or nonsuperconducting
origins of the PG. The resolution can be provided by a doping
phase diagram, both because Oxygen-doping is the most critical
HTSC parameter (HTSC can be altered from a metal to an
antiferromagnetic insulator by decreasing O-content) and because
distinctly different diagrams are expected for different scenarios
\cite{Tallon}. In a superconducting scenario, the PG represents
the pairing energy, which can be finite at $T > T_c$ in a strong
coupling case. The smaller gap represents the energy required for
maintenance of a long range coherence \cite{Deut} at $T < T_c$.
Those two energies should {\it merge} in the overdoped (OD)
region, as the weak coupling limit is approached. If, on the
contrary, the PG appears abruptly at some critical doping point
$p_c$ and the PG {\it crosses} the superconducting gap (SG) at the
phase diagram, it would correspond to a non-superconducting PG
\cite{Tallon}, which develops in the underdoped (UD) region at the
expense of the SG.

The present state of confusion requires further studies using
advanced experimental techniques. One of those is an interlayer
tunneling spectroscopy, which is unique in it's ability to measure
properties {\it inside} HTSC single crystals. This method is
specific to strongly anisotropic HTSC, like
Bi$_2$Sr$_2$CaCu$_2$O$_{8+\delta}$ (Bi2212), in which mobile
charge carriers are localized in double CuO$_2$ layers, while the
transverse ($c-$axis) transport is due to interlayer tunneling
\cite{Kleiner,Fiske}. Interlayer tunneling has become a powerful
tool for studying both electron\cite{KrTemp,KrMag,Suzuki} and
phonon\cite{Schlenga} DOS of HTSC. It has several important
advantages compared to surface tunneling techniques: (i) it probes
bulk properties and is insensitive to surface deterioration or
surface states\cite{Lanzara}; (ii) the current direction is well
defined; (iii) the tunnel barrier is atomically perfect and has no
extrinsic scattering centers; (iv) mesa structures are
mechanically stable and can sustain high bias in a wide range of
temperatures ($T$) and magnetic fields ($H$).

Here I present a systematic O-doping dependent interlayer
tunneling study of Bi2212. The spectroscopic resolution was
improved by decreasing in-plane mesa sizes, thus avoiding stacking
faults and self-heating in the mesas \cite{Heat}. This way it was
possible to trace the SG and the PG at $T \sim T_c$ and analyze
"closing" of the PG at a characteristic temperature $T^*$. The
obtained doping phase diagram exhibits a critical doping point for
appearance of the PG and a characteristic crossing of the SG and
the PG close to the optimal doping. This points towards
coexistence of two different, competing order parameters in HTSC.
Experimental data indicate that the SG can either form a combined
gap with the PG or remain uncombined at $T<T_c$ and that the
interlayer tunneling is predominantly coherent in OD samples, but
becomes progressively incoherent with decreasing doping.

Small mesa structures, with areas $A = 10-30\mu m^2$, containing
$N = 5 - 12$ intrinsic junctions, were made on top of Bi2212
single crystals by a microfabrication technique \cite{KrTemp}. The
fabrication was highly reproducible: all mesas on the same crystal
exhibited similar behavior, independent of $A$ and $N$. UD
crystals were prepared by annealing in vacuum at 600$^\circ C$.

Fig. 1 shows current-voltage characteristics (IVC's) per junction
at 4.2K for different doping. A characteristic knee in IVC's is
clearly seen, followed by a normal resistance branch $R_N$. The
knee is strongly suppressed both by $T$ \cite{KrTemp} and $H$
\cite{KrMag}, while $R_N$ is almost $T,H-$ independent. Such
behavior is typical for SIS-type tunnel junctions, in which the
knee occurs at a sum-gap voltage $2\Delta_{SG}/e$, where
$\Delta_{SG}$ is the maximum SG. Multiple branches at low bias
correspond to one-by-one switching

\begin{figure}
\noindent
\begin{minipage}{0.48\textwidth}
\epsfxsize=0.8\hsize \centerline{ \epsfbox{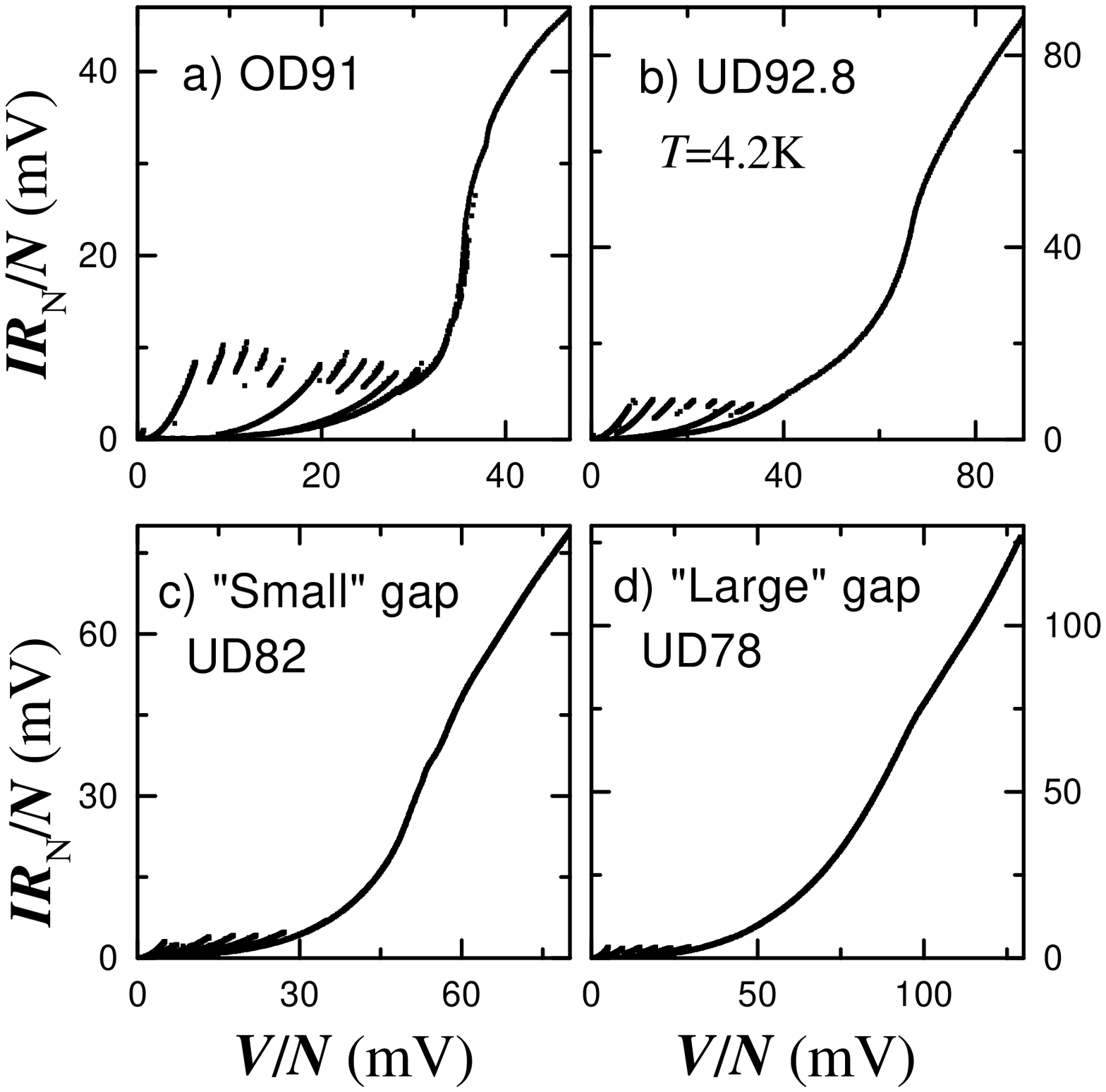} }
\vspace*{6pt} \caption{Normalized IVC's per junction at $T=4.2K$
for Bi2212 mesas with different O-doping: a) OD $T_c=91 K$, b)
slightly UD $T_c=92.8 K$, c) UD $T_c=82 K$ the small gap case; d)
UD $T_c=78 K$ the large gap case.}
\end{minipage}
\end{figure}

\noindent of junctions from a supercurrent to a quasiparticle (QP)
branch. QP branches carry important information: (i) the maximum
spacing between QP branches $\delta V_{QP}$ is an additional
parameter for estimation of the SG; (ii) the extent of QP branches
along the vertical axis in Fig. 1 represents the $I_c R_N$ product
per junction, which is a critical parameters of a Josephson
junction.

Figs. 2 and 3 show tunneling conductance $\sigma = dI/dV$ curves
for slightly OD and UD samples, respectively. Below $T_c$ a sharp
peak, corresponding to the knee in IVC's, is seen. The peak
voltage, $V_{peak}$, decreases as $T \rightarrow T_c$. Above $T_c$
the peak disappears, but a distinctly different dip-and-hump
structure remains, representing the persisting PG \cite{KrTemp}.
$T-$ dependencies of the peak (large symbols) and the hump (small
symbols + lines) voltages for four samples with different doping
are shown in Fig. 4.

For OD samples, $V_{peak}$ can be clearly traced up to $T_c$ and
$V_{peak} \rightarrow 0$ at $T_c$, see Figs. 2 a) and 4. Note that
$V_{peak}(T \ll T_c)$ is substantially larger than the hump
voltage $V_{hump}(T_c)$ in the OD mesa. At $\sim 150 K$,
$V_{hump}$ starts to decrease and vanishes at $T^* \sim 200 K$
(see Figs. 2 b) and 4). Interestingly, IVC's are nonlinear even
above $T^*$, see Fig. 2 b) and $\sigma(V)$ has an inverted
parabola shape, which might indicate the presence of van-Hove
singularity close to Fermi level in slightly OD samples
\cite{VHove}. Details of the PG closing at $T^*$ are important for
understanding the origin of the PG. At the first glance
$V_{hump}(T \rightarrow T^*)$ resembles a BCS-like dependence,
typical for a phase transition due to an onset of charge or spin
density waves~\cite{CSDW}. However, a different perspective opens
when the parabolic background at $T>T^*$ is subtracted, see Fig. 2
c). In such a plot the PG simply "fills-in" at $T^* \simeq
\Delta_{PG}$ without a significant change in $V_{hump}$. This may
indicate that there is a smooth

\begin{figure}
\noindent
\begin{minipage}{0.48\textwidth}
\epsfxsize=0.75\hsize \centerline{ \epsfbox{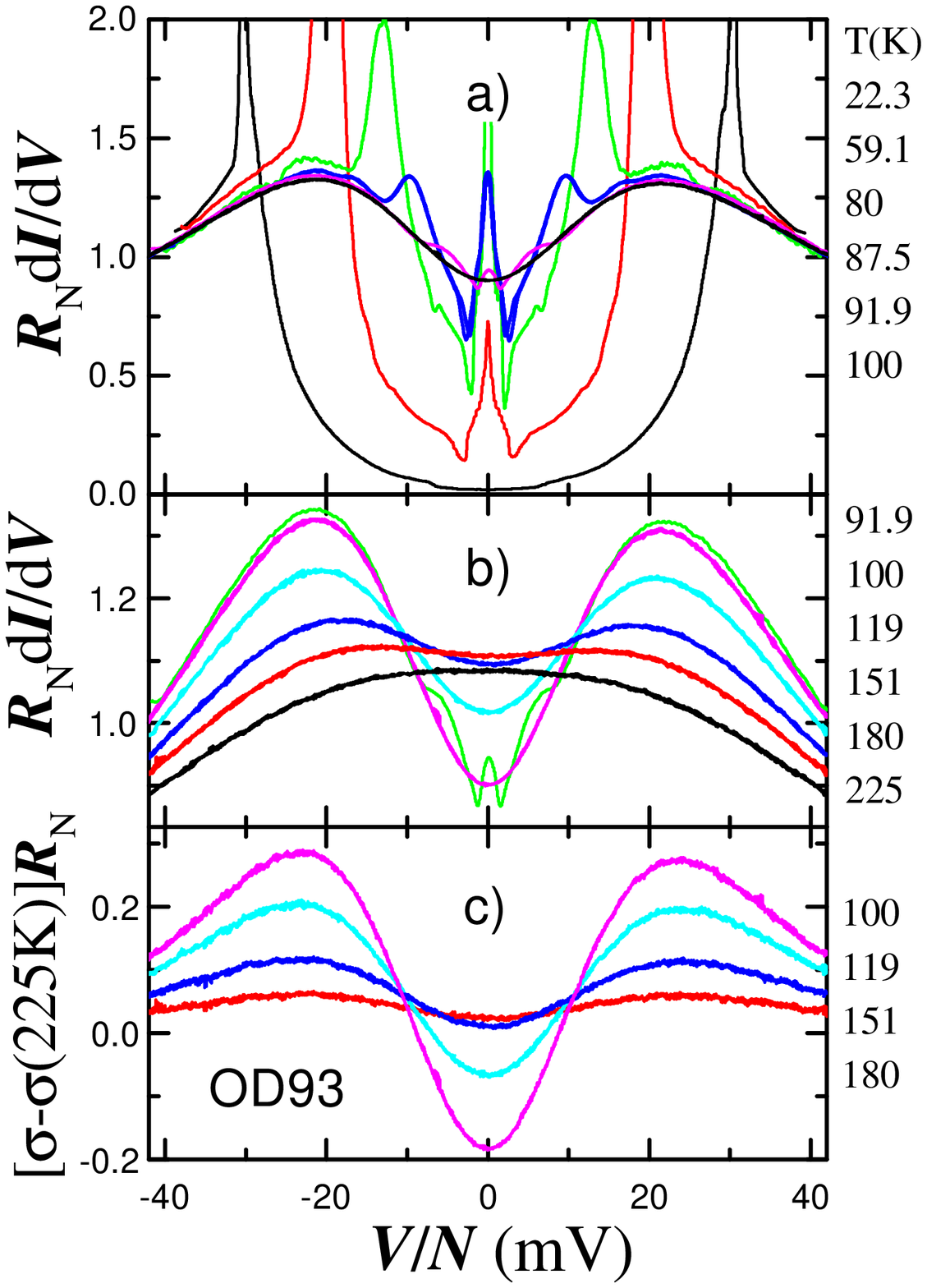} }
\vspace*{6pt} \caption{ d$I$/d$V$($V$) curves for a slightly OD
sample $T_c = $ 93 K: a) below and just above $T_c$; b) just below
and above $T_c$; c) with a subtracted parabolic background.}
\end{minipage}
\end{figure}

\noindent crossover rather than a true phase transition at $T^*$.

The behaviour of the SG in UD samples at $T \rightarrow T_c$ is
one of the most important and yet controversial issues
\cite{KrTemp,Fisher}. For UD samples the peak is much weaker than
for OD samples even at low $T$, cf. Figs. 2 a) and 3 a), and it
rapidly smears out with increasing $T$. The contrast of the small
peak can be increased by subtracting the background PG
dip-and-hump at $T > T_c$, as shown in Fig. 3 c). Thus $V_{peak}$
can be located at $T \sim T_c$.

UD samples showed two distinct types of behavior, which I refer to
as "small" and "large" gap cases, cf. Figs. 1 c) and d). d$I$/d$V$
curves for the large gap case are shown in Fig. 3. The behavior of
large and small gaps is different: (i) for large gaps, the
dip-and-hump is strongly enhanced at the expense of the peak,
e.g., in Fig. 3 the PG dip-and-hump is clearly recognizable even
at low $T$. For the large (UD84.4) and small (UD85) gap samples,
in Fig. 4, ratios of hump to dip conductances
$\sigma(V_{hump})/\sigma(0)$ at 100 K are $\sim$ 5.2 and 1.8,
while $\sigma(V_{peak}) R_N$ at 4.2K is $\sim$ 1.6 and 10,
respectively. (ii) For small gaps $V_{peak} \rightarrow 0$ at
$T_c$ and decreases with UD together with $T_c$, while for large
gaps $V_{peak}$ remains finite at $T_c$ (even though it drops
considerably at $T_c$) see Figs. 3 c) and 4, and both peak and
hump voltages increase with underdoping despite the decrease of
$T_c$. (iii) Noticeably, apart from gap magnitudes, other
parameters, such as $\rho_c$, $J_c$ and $I_c R_N$ are similar,
implying that the tunneling barrier is not affected.

The observed differences can be explained by the following
scenarios for formation of small and large gaps,

\begin{figure}
\noindent
\begin{minipage}{0.48\textwidth}
\epsfxsize=0.8\hsize \centerline{ \epsfbox{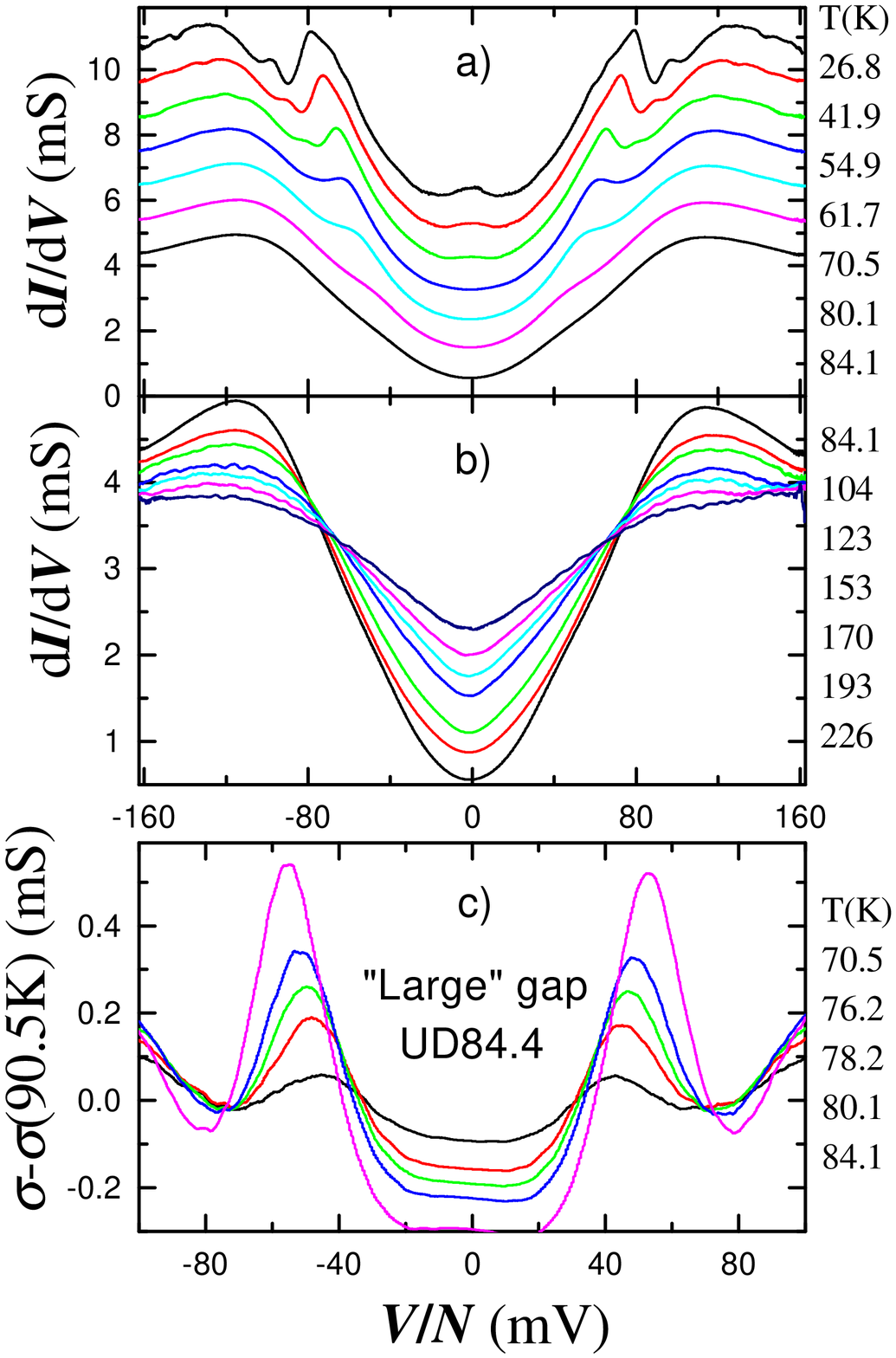} }
\vspace*{6pt} \caption{d$I$/d$V$ curves for the large gap case, UD
sample $T_c=84.4 K$: a) below $T_c$ (curves are shifted for
clarity), b) above $T_c$ and c) curves with a subtracted PG
background in the vicinity of $T_c$. Note different scales in
Figs. a,b)and c). }
\end{minipage}
\end{figure}

\noindent shown schematically in insets a) and b) of Fig. 4:

The small gap is developed on top of a modest suppression of the
DOS at Fermi level, i.e., when there is no true gap at $T_c$,
which might interfere with the opening SG. Therefore, the peak in
d$I$/d$V$ represents the bare (uncombined) SG, which vanishes at
$T_c$, while the dip-and-hump represent a "normal" background,
which is hindered by the growing SG. Such behavior was observed
for OD, optimally doped\cite{KrTemp} and UD samples with the small
gap, see Figs. 2 a) and 4.

On the other hand, in the large gap case the SG is developed on
top of a true gap $\Delta_0$, see inset b) in Fig. 4. Indeed, from
Fig. 3 b) it is seen that the PG dip-and-hump flatten with
increasing $T$ in a state conserving manner, characteristic for a
"true" energy gap in DOS, and $\sigma(V)$ curves intersect in one
point, indicating approximately constant value of the PG in the
measured $T-$ range. Below $T_c$ this causes formation of the
combined ($\Delta_0$ and $\Delta_{SG}$) large gap. In agreement
with this assumption: (i) the large gap does not vanish, but
approaches $\Delta_0$ at $T_c$, see Figs. 3a) and 4. (ii) The peak
completely disappears at $T_c$ but does not transform into the
hump because $eV_{hump} > \Delta_0$, see the dashed line in inset
b). (iii) The volume of the peak (superfluid density) is small
because it builds up from an initially suppressed DOS. (iv) The
opening of the SG at $T < T_c$ shifts all DOS features, including
the hump, as shown in inset b)

\begin{figure}
\noindent
\begin{minipage}{0.48\textwidth}
\epsfxsize=0.8\hsize \centerline{ \epsfbox{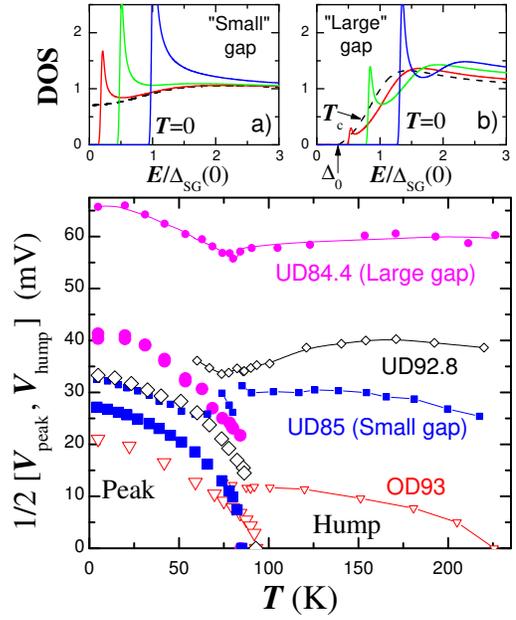} }
\vspace*{6pt} \caption{$T-$ dependencies of $\frac12 V_{peak}$
($\sim$ SG, large symbols) and $\frac12 V_{hump}$ ($\sim$ PG,
small symbols + lines) for slightly OD (triangles) and UD
(diamonds) samples and for UD samples with small (squares) and
large (circles) gaps. Insets a,b) show scenarios for formation of
small and large gaps at four different temperatures from 0 to
$T_c$ (dashed lines). }
\end{minipage}
\end{figure}

\noindent of Fig. 4. The {\it correlated} shift of both the peak
and the hump with $T$ for UD84.4 sample, as shown in Figs. 3 and
4, is a strong argument in favor of the combined scenario of the
large gap. Similarly, {\it uncorrelated} $T-$ dependent peak and
$T-$ independent hump in the OD93 sample, see Figs. 2 and 4,
suggests that the small gap represents the uncombined SG.
Interestingly, if we take $V_{peak}(4.2K)-V_{peak}(T_c)$ as a
measure of the SG part of the combined gap, it will coincide with
the small gap for a similar doping, as shown by arrows in Fig. 5
d). A systematic increase of the hump energy with decreasing $T$,
observed by ARPES \cite{Campuz}, would have been consistent with
the combined scenario of the large gap if not for the lack of
correlated $T-$dependence of the coherence peak.

Fig. 5 shows O-doping dependencies of: a) $T_c$, dashed line
represents the empirical expression, used for estimation of $p$;
b) the critical current density, $J_c$, and the $I_c R_N$ product
per junction; c) the tunneling resistivity at large bias $\rho_c =
R_N A/(N s)$. The $I_c R_N$ is an important parameters of a
Josephson junction. As Bi2212 is likely to be a d-wave
superconductor\cite{Tsue}, the $I_c R_N$ depends both on $\Delta
_{SG}$ and the coherence (in-plane momentum conservation) of
$c-$axis tunneling (another highly debated issue in
HTSC\cite{PWA}). The $I_c R_N$ is maximum $\simeq \Delta_{SG}/e$
for coherent, and zero for completely incoherent
tunneling\cite{dwave}. For OD mesas $I_c R_N \sim 10$mV is a
considerable fraction $\sim 0.6$ of $\Delta_{SG}/e$, indicating
predominantly coherent nature of the interlayer tunneling. With
underdoping,

\begin{figure}
\noindent
\begin{minipage}{0.48\textwidth}
\epsfxsize=0.85\hsize \centerline{ \epsfbox{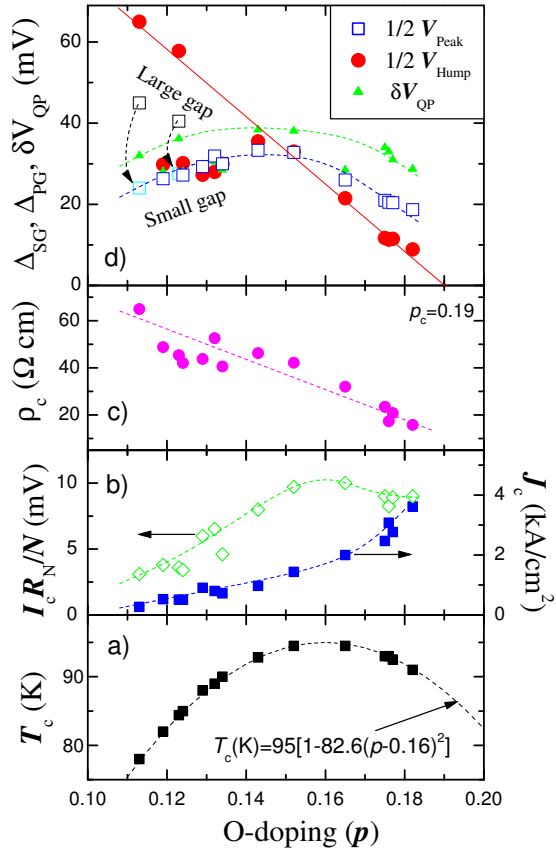} }
\vspace*{6pt} \caption{Doping dependencies of: a) $T_c$; b) $J_c$
and $I_c R_N$. It is seen that $I_c R_N$ decreases dramatically
with doping, due to progressively incoherent nature of the
interlayer tunneling. c) $\rho_c$ at large bias; d) the doping
phase diagram of Bi2212: $\frac12 V_{peak}(4.2K)$ (open squares),
$\frac12 V_{hump}(100K)$ (solid circles), $\delta V_{QP}(4.2K)$
(triangles). A characteristic crossing of the SG and the PG and
the existence of critical doping point, $p_c \simeq 0.19$, are
clearly seen. }
\end{minipage}
\end{figure}

\noindent the $I_c R_N$ decreases dramatically at a much faster
rate than $\Delta_{SG}$. This indicates that the interlayer
tunneling becomes progressively incoherent in UD Bi2212.

Fig. 5 d) shows the obtained doping phase diagram of Bi2212. Here
I plot $\frac12 V_{peak}(4.2K)$ $\sim \Delta _{SG}/e$, $\frac12
V_{hump}(100 K > T_c)$ $\sim \Delta _{PG}/e$ and $\delta V_{QP}
(4.2K)$,
It is seen that the small gap ($\sim$ SG) shows a
similar tendency as $T_c$ and decreases both on OD and UD sides.
This is also supported by a correlated behavior of $\delta
V_{QP}$. In contrast, the large gap ($\sim$ PG) increases
approximately linearly with underdoping, as shown by the solid
line. The PG and the SG lines cross at about the optimal doping,
$p=0.16$. On the OD side the PG becomes considerably less than the
SG and shows a clear tendency to vanish at the critical doping
point, $p_c \simeq 0.19$. This speaks in favor of a
nonsuperconducting origin of the PG \cite{Tallon}, consistent with
earlier observations of different $T$ \cite{KrTemp} and $H$
\cite{KrMag} dependencies of the SG and the PG. Within such a
scenario, a suppression of superconductivity (decrease of $T_c$,
$\Delta_{SG}$, the superfluid density, etc.) in UD HTSC is caused
by appearance of the competing order parameter (PG), e.g., due to
strengthening of antiferromagnetic correlations and formation of
spin density waves. Note that a similar phase diagram, attributed
to competition between superconducting and antiferromagnetic
orders, was reported for heavy fermion
superconductors\cite{Mathur}.

At present, the reason for appearance of either small or large
gaps in UD samples is unclear. However, it is not due to
irreproducibility of fabrication (all mesas on the same crystal
show the same behavior) or macroscopic defects (regular QP
branches are observed in both cases). Presumably, the ambiguity is
connected with a microscopic inhomoginiety of UD
crystals\cite{Sacks}. The presence of ambiguity obscures
identification of the genuine HTSC behavior in the UD region.
However, there is no ambiguity for overdoped and optimally doped
samples. Therefore, conclusions that there is a critical doping
point in HTSC phase diagram and that the SG and the PG cross
rather than merge near the optimal doping are robust.

In summary, O-doping dependence of Bi2212 was studied using high
resolution interlayer tunneling spectroscopy. We were able to
simultaneously trace the superconducting gap and the $c-$axis
pseudo-gap at $T \sim T_c$ and analyze "closing" of the PG at
$T^*$. The obtained doping phase diagram exhibits a critical
doping point for appearance of the PG and a characteristic
crossing of the SG and the PG close to the optimal doping,
indicating a competing nature of two coexisting order parameters
in HTSC. In UD samples, the SG can either form a combined gap with
the PG or remain uncombined at $T<T_c$, but the bare SG vanishes
at $T \simeq T_c$ for all studied doping levels. Analysis of $I_c
R_N$ vs. $\Delta_{SG}$ indicates that the interlayer tunneling is
predominantly coherent in OD, but becomes progressively incoherent
in UD samples.

\end{multicols}
\end{document}